\documentclass[10pt,aps,prd,twocolumn,floats,floatfix,showpacs,superscriptaddress,nofootinbib]{revtex4-1}
\usepackage[utf8]{inputenc}               		

\usepackage{graphicx,mathtools,amssymb,amsmath,amsthm,amsfonts,epsfig,epsf}
\usepackage[outdir=./]{epstopdf}
\usepackage[linktocpage]{hyperref}
\hypersetup{pdfstartview=}
\usepackage[usenames]{color}		
\usepackage{tensor}	
\usepackage{csquotes}
\usepackage{mathrsfs}
\usepackage[caption=false]{subfig}

\newcommand{\dd}{\mathrm{d}}

\begin{document}

\title{The onset of spontaneous scalarization in generalised scalar-tensor theories}

\author{Giulia Ventagli}
\affiliation{School of Mathematical Sciences, University of Nottingham,
University Park, Nottingham NG7 2RD, United Kingdom}

\author{Antoine Lehébel}
\affiliation{School of Mathematical Sciences, University of Nottingham,
University Park, Nottingham NG7 2RD, United Kingdom}
\affiliation{School of Physics and Astronomy, University of Nottingham,
University Park, Nottingham NG7 2RD, United Kingdom}

\author{Thomas P.~Sotiriou}
\affiliation{School of Mathematical Sciences, University of Nottingham,
University Park, Nottingham NG7 2RD, United Kingdom}
\affiliation{School of Physics and Astronomy, University of Nottingham,
University Park, Nottingham NG7 2RD, United Kingdom}


\begin{abstract}
In gravity theories that exhibit spontaneous scalarization, astrophysical objects are identical to their general relativistic counterpart until they reach a certain threshold in compactness or curvature. Beyond this threshold, they acquire a non-trivial scalar configuration, which also affects their structure. The onset of scalarization is controlled only by terms that contribute to linear perturbation around solutions of general relativity. The complete set of these terms has been identified for generalized scalar-tensor theories. Stepping on this result, we study the onset on scalarization in generalized scalar-tensor theories and determine the relevant thresholds in terms of the contributing coupling constants and the properties of the compact object. 
\end{abstract}

\maketitle

\section{Introduction}

Gravitational wave astronomy now offers a way to observe the inspiral and coalescence of pairs of compact astrophysical objects, either black holes or neutron stars \cite{TheLIGOScientific:2017qsa,LIGOScientific:2018mvr}. This sheds light on two unexplored corners of gravity. First, during the late stages of the inspiral, the objects move at speeds close to the speed of light, allowing to test the dynamical regime of gravity. Second, the coalescence of compact objects involves high curvatures and non-linearity plays a significant role. Strong field effects can emerge progressively as the curvature increases. They can also appear abruptly, in a phase transition process. The study of these phase transitions has a rich history, particularly in the context of scalar-tensor gravity. It was originally investigated by Damour and Esposito-Far\`ese (DEF) \cite{Damour:1993hw} in the context of neutron stars, and later dubbed spontaneous scalarization (in analogy with magnetization).

The DEF model relies on a specific coupling between the scalar field and the metric. The model admits several branches of solutions, among which all general relativity (GR) solutions with a trivial scalar field. At low curvatures, GR solutions are stable against scalar perturbations. However, in the strong field regime, the scalar field acquires a tachyonic effective mass that destabilizes the solution \cite{Harada:1997mr}. This instability is ultimately quenched by non-linear terms, resulting in a neutron star with a non-trivial scalar profile \cite{Novak:1998rk}. The DEF model was studied thoroughly, especially in the light of binary pulsar constraints (e.g., \cite{Freire:2012mg,Antoniadis:2013pzd,Shao:2017gwu}). These constraints have severely constrained the model in its original formulation, but they can be straightforwardly evaded by giving the scalar field a bare mass \cite{Ramazanoglu:2016kul}.

It has recently been shown that the DEF coupling is not unique in generating scalarization. In particular, scalar-Gauss-Bonnet gravity exhibits similar effects, both for neutron stars and black holes \cite{Silva:2017uqg,Doneva:2017bvd}. A generic property of these models is that linear terms in the field equations determine the onset of scalarization, while non-linear terms keep the instability under control and determine the endpoint of the process \cite{Blazquez-Salcedo:2018jnn,Silva:2018qhn,Macedo:2019sem}. It is also possible to extend the concepts of scalarization to different field contents (e.g., \cite{Herdeiro:2018wub,Ramazanoglu:2017xbl,Ramazanoglu:2018hwk}).
At this point, it is legitimate to wonder what is the most generic model that can trigger scalarization. Within the context of Horndeski theory, this question was answered in \cite{Andreou:2019ikc}. Horndeski theory offers a robust framework of models which retain second-order field equations, and are therefore free of Ostrogradsky ghost \cite{Horndeski:1974wa}. Reference \cite{Andreou:2019ikc} listed all the terms that can play a role in the onset of scalarization. In this paper, we determine quantitatively the bounds on the couplings that allow scalarization to take place. We combine the effect of all relevant couplings simultaneously and also examine how the structure of the compact object affects the threshold of the instability.

The paper is organized as follows. In sec.~\ref{sec:setup}, we describe the framework that allows us to identify the onset of an instability. Then, in sec.~\ref{sec:effmass}, we investigate the effect of the parameters that can generate an effective mass. Section~\ref{sec:star} is devoted to the influence of the background on the instability (mass of the object, equation of state). In sec.~\ref{sec:a41}, we focus on the parameter that deforms the effective metric in which perturbations propagate. Finally, we conclude in sec.~\ref{sec:conc}.

\section{Setup}
\label{sec:setup}

As mentioned above, in the framework of Horndeski theory, all the terms that can trigger a tachyonic instability (or play a role in its onset) were listed in~\cite{Andreou:2019ikc}. The minimal action containing all these terms reads
\begin{equation}
\begin{split}\label{eq:ActionCaseI}
S&=\int\dd^4x\sqrt{-g}\bigg\{\dfrac{R}{2\kappa}+X+ \gamma\, G^{\mu\nu}\nabla_\mu\phi\,\nabla_\nu\phi
\\
&\quad -\left(m_\phi^2+\dfrac{\beta}{2} R-\alpha\mathscr{G}\right)\dfrac{\phi^2}{2}\bigg\}
 +S_\mathrm{M},
\end{split}
\end{equation}
where $X=-\nabla_\mu\phi\nabla^\mu\phi/2$, $\kappa=8\pi G/c^4$ and $\mathscr{G}$ is the Gauss-Bonnet invariant:
\begin{equation}
\mathscr{G}=R^2-4R_{\mu\nu}R^{\mu\nu}+R_{\mu\nu\rho\sigma}R^{\mu\nu\rho\sigma}.
\end{equation}
$m_\phi$ is the bare mass of the scalar field, while $\gamma$, $\beta$ and $\alpha$ parametrize deviations with respect to GR; $\beta$ is dimensionless, while $\alpha$ and $\gamma$ have the dimension of a length squared. $S_\mathrm{M}$ denotes the matter action, where matter is assumed to couple minimally to the metric only: we are working in the so-called Jordan frame. Note that $\beta$ is defined in order to match the notation used in the case of the (linearized) DEF model. The precise relation to the original DEF model is discussed in detail in Ref.~\cite{Andreou:2019ikc}. In general, the action in eq.~\eqref{eq:ActionCaseI} admits all GR solutions with a trivial scalar field configuration $\phi=0$. One can obtain from this action any theory within the Horndeski class that admits GR solution with a constant scalar (not necessarily zero) and exhibits spontaneous scalarization: first one can shift the scalar by a suitable constant and then supplement the action with the desired nonlinear interactions.

Our goal will be to investigate whether GR solutions with $\phi=0$ are stable or not, by focusing on the perturbations of the scalar field. The scalar field equation associated with the action \eqref{eq:ActionCaseI} reads
\begin{equation}
\label{LinEq}
\tilde{g}^{\mu\nu}\nabla_\mu\nabla_\nu\phi-m^2\phi=0,
\end{equation}
where the effective metric and the mass term are respectively 
\begin{align}
\label{effMetr}
\tilde{g}^{\mu\nu}&=g^{\mu\nu} - \gamma G^{\mu\nu},
\\
\label{massI}
m^2&=m_\phi^2+\frac{\beta}{2} R-\alpha\mathscr{G}.
\end{align}
It is clear from these equations that $\beta$ and $\alpha$ generate an effective mass for $\phi$ in a curved background. On the other hand, $\gamma$ determines the effective metric that defines the d'Alembertian which acts on the scalar field perturbations. Equation \eqref{LinEq} is linear by construction, since we kept in the action only the terms that contribute linearly. This approach is valid when focusing on the onset of the scalarization, when linear terms dominate. In order to determine the final state of the process, one needs to include non-linear terms as well. Additionally, we work in the decoupling limit where we only perturb the scalar field. These perturbations will eventually back-react onto the metric, and a consistent analysis (beyond the onset of the instability) should thus include metric perturbations. 

GR vacuum solutions are Ricci flat. Therefore, it is immediate from action \eqref{eq:ActionCaseI} that only the Gauss-Bonnet coupling $\alpha$ controls the scalarization of vacuum solutions (in particular electrically neutral black holes). Since we are interested in the combined effect of the parameters involved in \eqref{eq:ActionCaseI}, we focus on neutron stars, where matter is present under the form of a perfect fluid. We further consider a static and spherically symmetric background spacetime:
\begin{equation}
\text{d}s^2=-h(r)c^2\text{d}t^2+f(r)^{-1}\text{d}r^2+r^2\text{d}\Omega^2.
\label{eq:gansatz}
\end{equation}
The metric functions $h$ and $f$ are determined as solutions of an equivalent to the Tolman-Oppenheimer-Volkoff system of equations~\cite{Tolman:1939jz,Oppenheimer:1939ne}, together with the pressure $P$ and energy density $\epsilon$ of the perfect fluid that composes the star. These equations can be found in appendix \ref{sec:TOV}. In this framework, one has to specify some equation of state $P(\epsilon)$. We use two equations of state, SLy and MPA1~\cite{Gungor:2011vq}, both favored by LIGO-Virgo tidal measurements~\cite{TheLIGOScientific:2017qsa}, which seem to prefer soft equations of state. We work in units where $c=1$, $G=1$ and $M_\odot=1$.

Thanks to spherical symmetry, we can decompose the scalar perturbation on the basis of spherical harmonics:
\begin{equation}
\phi=\sum_{\ell,m}\hat\phi_{\ell m}(t,r)Y_{\ell m}(\theta,\phi).
\end{equation}
We will focus on the breather mode, $\ell=m=0$, which is the first one to exhibit instability when it is present. In order to make the scalar field equation more transparent, we rescale this mode according to $\hat\phi_{00}(t,r) =K(r)\sigma(t,r)$ with 
\begin{equation}
\begin{split}
K(r)&=\left\{ r^2 - 2\gamma\left[-1+ \dfrac{f}{h} (rh)'\right]\right\}^{-1/4}
\\
&\quad\times\{r^2 -2 \gamma[ (rf)'-1]\}^{-1/4},
\end{split}
\end{equation}
and we trade off the radial coordinate $r$ for a new one, $r_\ast$, defined through
\begin{equation}
\frac{dr_\star}{dr}=\frac{\sqrt{h}}{\sqrt{f} K^2 \left[ 2 \gamma r f h'-(2 \gamma+r^2-2 \gamma f)h \right]}\,.
\label{eq:drast}
\end{equation}
Equation \eqref{LinEq} then takes the following form:
\begin{equation}\label{scalarEqPot}
-\dfrac{1}{c^2}\,\frac{\partial^2\sigma}{\partial t^2}+\frac{\partial^2\sigma}{\partial r_\ast^2}=V_{\text{eff}}(r_\ast)\,\sigma,
\end{equation} 
where $V_\text{eff}$ depends on the parameters of the action \eqref{eq:ActionCaseI} as well as on the background geometry. Its full expression can be found in appendix~\ref{App:Veff}. We further focus on exponentially growing perturbations: $\sigma(t,r_\ast)=\hat\sigma(r_\ast) e^{\omega t}$, with $\omega>0$.\footnote{One could also look for the quasi-normal modes associated with $V_\text{eff}$, allowing complex values of $\omega$. However, this requires a much wider set-up, which is not needed to establish the presence of an instability.} Equation \eqref{scalarEqPot} then boils down to a Schr\"odinger equation:
\begin{equation}\label{scalarEqPot2}
\frac{\mathrm{d}^2\hat\sigma}{\mathrm{d} r_\ast^2}=\left[V_{\text{eff}}(r_\ast)+\left(\dfrac{\omega}{c}\right)^2\right]\hat\sigma,
\end{equation} 
where $V_\mathrm{eff}$ is clearly an effective potential, and $-(\omega/c)^2$ plays the role of the energy of the perturbation. The existence of a bound state for $V_\mathrm{eff}$ with `energy' $E_0<0$ implies the existence of an instability, with characteristic growth rate $\omega=c\sqrt{-E_0}$. Our strategy will thus be the following: we start from values of the parameters for which the theory reduces to GR with a minimally coupled scalar field and hence there cannot be any instability. We gradually increase the parameters, thus progressively deforming the potential. Whenever a bound state appears for $V_\mathrm{eff}$, we identify it with a new unstable mode for $\phi$. By continuity, when deforming the potential, a new bound state will appear with a vanishing energy, $E_0=0$. Therefore, we solve eq.~\eqref{scalarEqPot2} for $\omega=0$, while scanning the parameters $\beta$, $\alpha$ and $\gamma$. This is less intuitive than choosing a set of parameters and scanning $\omega$, but the final result is equivalent, and the procedure is easier to implement. 

Equation \eqref{scalarEqPot2} will admit a solution for any set of values of the parameters. Among these solutions, we identify as a bound state those with $\mathrm{d}\hat\sigma/\mathrm{d}r_\ast\to0$ for $r_\ast\to+\infty$.\footnote{On a more technical level, we perform a numerical integration of eq.~\eqref{LinEq} expressed in terms of $r$, rather than in terms of $r_\ast$ as in eq.~\eqref{scalarEqPot2}; we extract $\mathrm{d}\sigma/\mathrm{d}r$ at a radius $r_\mathrm{max}$ equal to 200 times the Schwarzschild radius of the star.} Physically, this is necessary for the scalar perturbation to be localized in space. In terms of quantum mechanics, this corresponds to the fact that $K(r_\ast)\hat\sigma(r_\ast)$ has to be square integrable. Note that, contrary to the naive expectation from eq.~\eqref{scalarEqPot2}, it is $K(r_\ast)\hat\sigma(r_\ast)$ that should be interpreted as the wave function, rather than $\hat\sigma(r_\ast)$ itself; this is very similar to the $1/r$ rescaling of the wave functions that allows to solve for the bound states of 3D spherically symmetric quantum wells (see e.g., secs.~14-16 of~\cite{schiff1955quantum}).

\section{Changing the effective mass}
\label{sec:effmass}

As mentioned above, the main terms that can contribute to the effective mass $m^2$ are the bare mass term of the scalar field, the coupling between $\phi$ and the Ricci scalar, and the coupling between $\phi$ and the Gauss-Bonnet invariant. These terms are parametrized by three constants: $m_\phi$, $\beta$ and $\alpha$.
Although the terms proportional to $\gamma$ can affect the instability threshold, they contribute only as a multiplicative constant. Therefore, we will set $\gamma=0$ throughout this section, and explore the role of this parameter in full detail in Sec.~\ref{sec:a41}.
Several works already investigated the influence of each of the parameters $m_\phi$, $\beta$ and $\alpha$ separately ({\em e.g.},~\cite{Damour:1993hw,Harada:1997mr,Silva:2017uqg,Doneva:2017bvd,Antoniou:2017acq,Blazquez-Salcedo:2018jnn,Minamitsuji:2018xde,Silva:2018qhn,Macedo:2019sem,Doneva:2019vuh}). We study how varying several parameters simultaneously affects the threshold; in this way, we explore much wider regions of the parameter space. Most of our results are presented as 2D plots, where we freeze all parameters but two, and show the stable/unstable regions.

\subsection{Coupling to curvature invariants}
\label{sec:alphabeta}

\begin{figure}
\subfloat{\includegraphics[height=0.347\textwidth, width=0.1\textwidth]{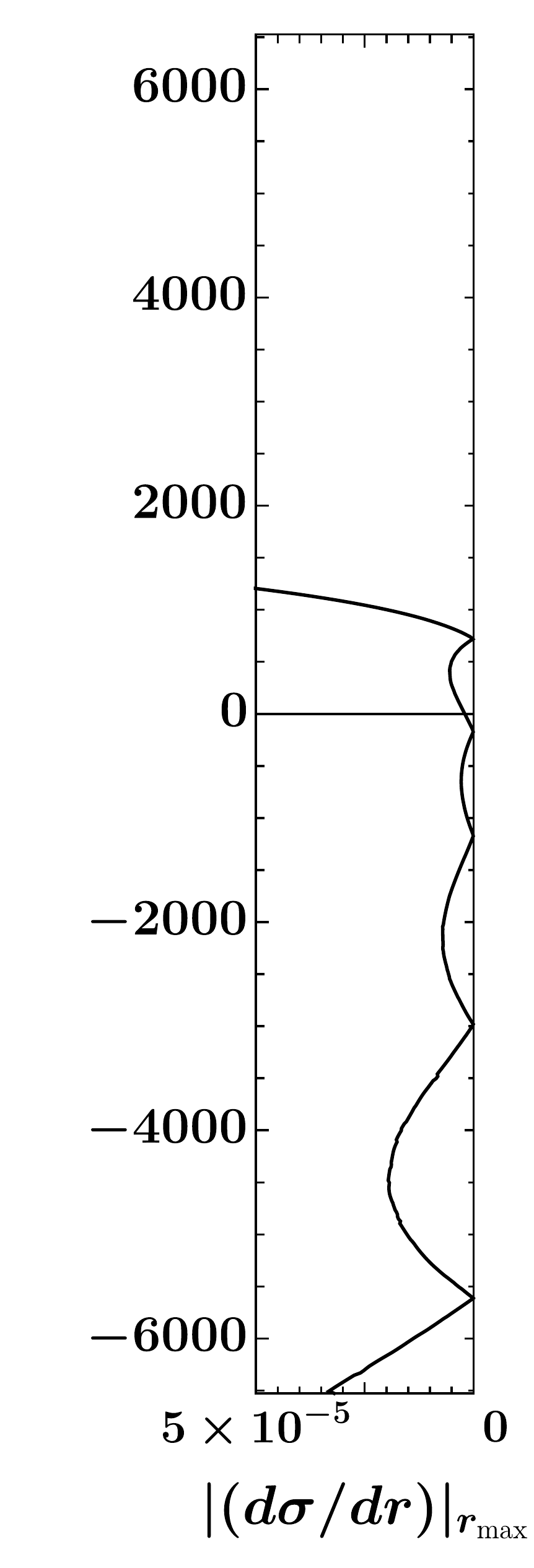}%
\quad\includegraphics[height=0.35\textwidth, width=0.35\textwidth]{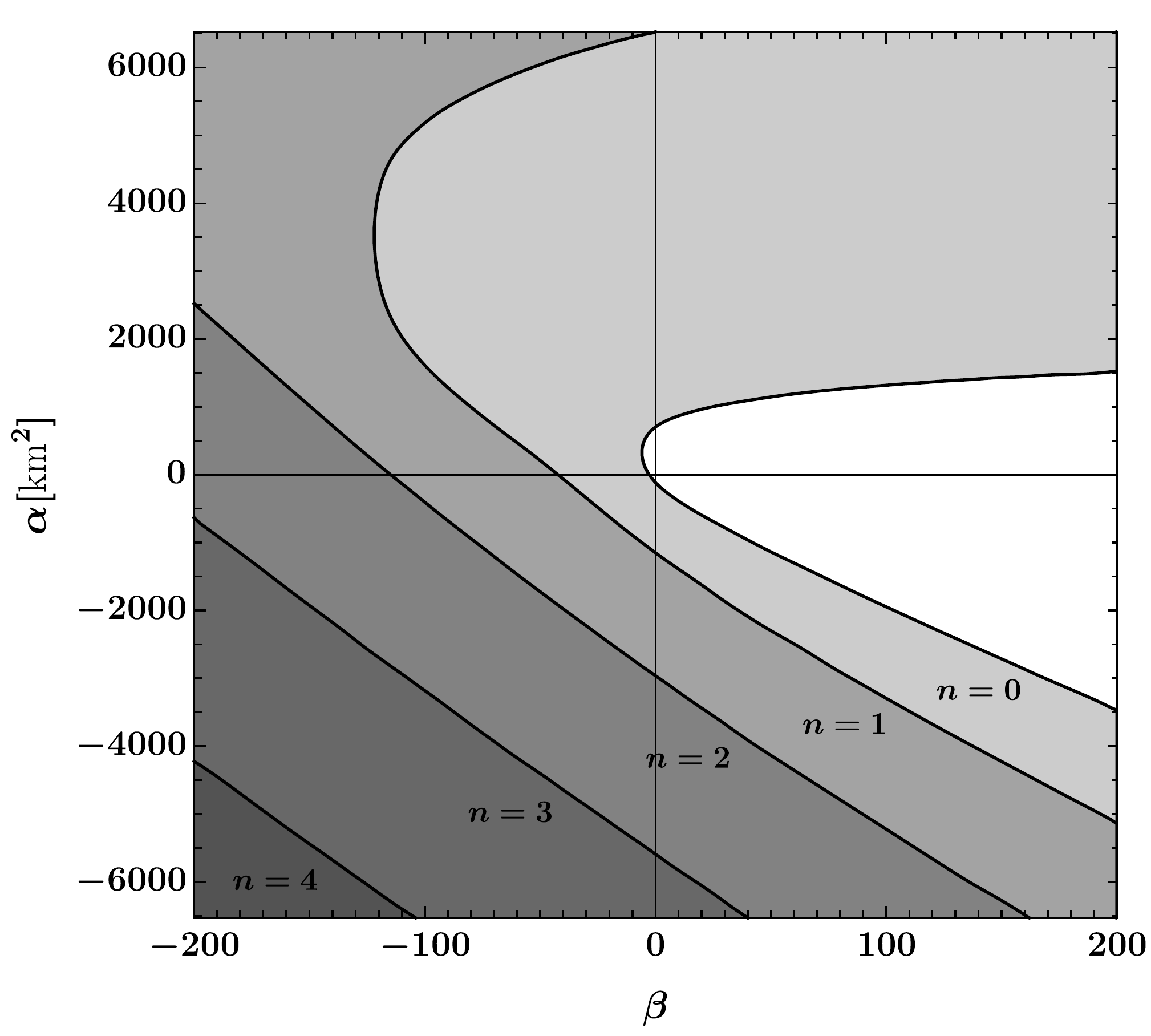}}
\\
\flushright
\subfloat{\includegraphics[width=0.368\textwidth]{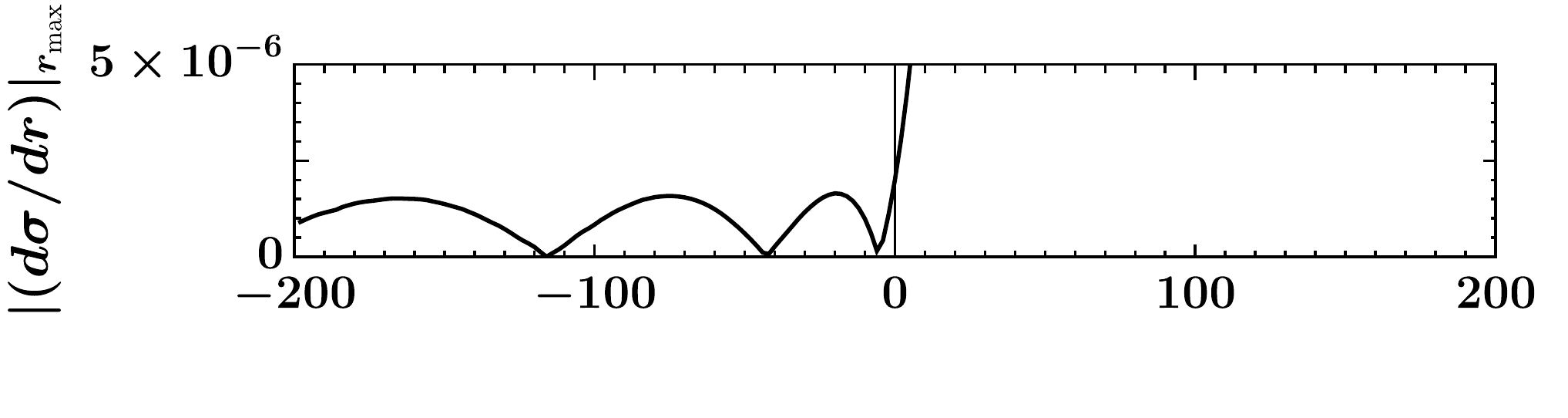}}
\caption{Stable and unstable regions in the $(\beta,\alpha)$ space for a light star ($M=1.12~M_\odot$, SLy equation of state). In the 2D plot, each line is labeled according to the number of nodes $n$ of the corresponding unstable mode. Inside the white region, where the point $(\beta,\alpha)=(0,0)$ lies, the GR solution is stable. Every line crossed while moving away from the origin corresponds to the appearance of a new unstable mode; any point in parameter space that lies within a grey region corresponds to an unstable solution. The lower panel shows $|\mathrm{d}\sigma/\mathrm{d}r|$ at $r_\mathrm{max}$ when varying $\beta$ in the same range as the 2D plot, with $\alpha=0$; it can be understood as a cut in the $(\beta,\alpha)$ plane along the $\beta$ axis (each cusp corresponding to a line-crossing in the 2D plot). Similarly, the left panel shows a cut along the $\alpha$ axis.}
\label{fig:alphabeta}
\end{figure}

We first consider a vanishing bare mass, $m_\phi=0$. The model is then parametrized by $\beta$ and $\alpha$ only. The background we consider is a neutron star described by the SLy equation of state. We choose its central energy density to be $\rho_\mathrm{c}=8.1\times10^{17}$~kg/m$^3$, so that its gravitational mass is $M=1.12~M_\odot$, which corresponds to the bottom of the mass range for observed neutron stars~\cite{Martinez:2015mya,Suwa:2018uni}. The radius of the star is then $R_\mathrm{s}=11.7$~km. The results are summarized in Fig.~\ref{fig:alphabeta}.
The white area corresponds to the region of the parameter space where the background solution is stable. A new unstable mode appears when crossing each line while moving away from the origin. The lines are labeled with the number of nodes $n$ of the associated mode, ranging from 0 to infinity. The $n=0$ line is the boundary of the stable region. Any choice of parameters beyond this line will make the GR solution unstable. The left and bottom panels shows cuts in the $(\beta,\alpha)$ plane, along the $\alpha$ and $\beta$ axes respectively; these panels actually reproduce known results, e.g. of~\cite{Harada:1997mr} and~\cite{Silva:2017uqg}. Notably, the scalarization threshold when only $\beta$ is presents takes the well-known order of magnitude, $\beta=-5.42$.

\begin{figure}
\subfloat[]{%
  \includegraphics[width=0.49\linewidth]{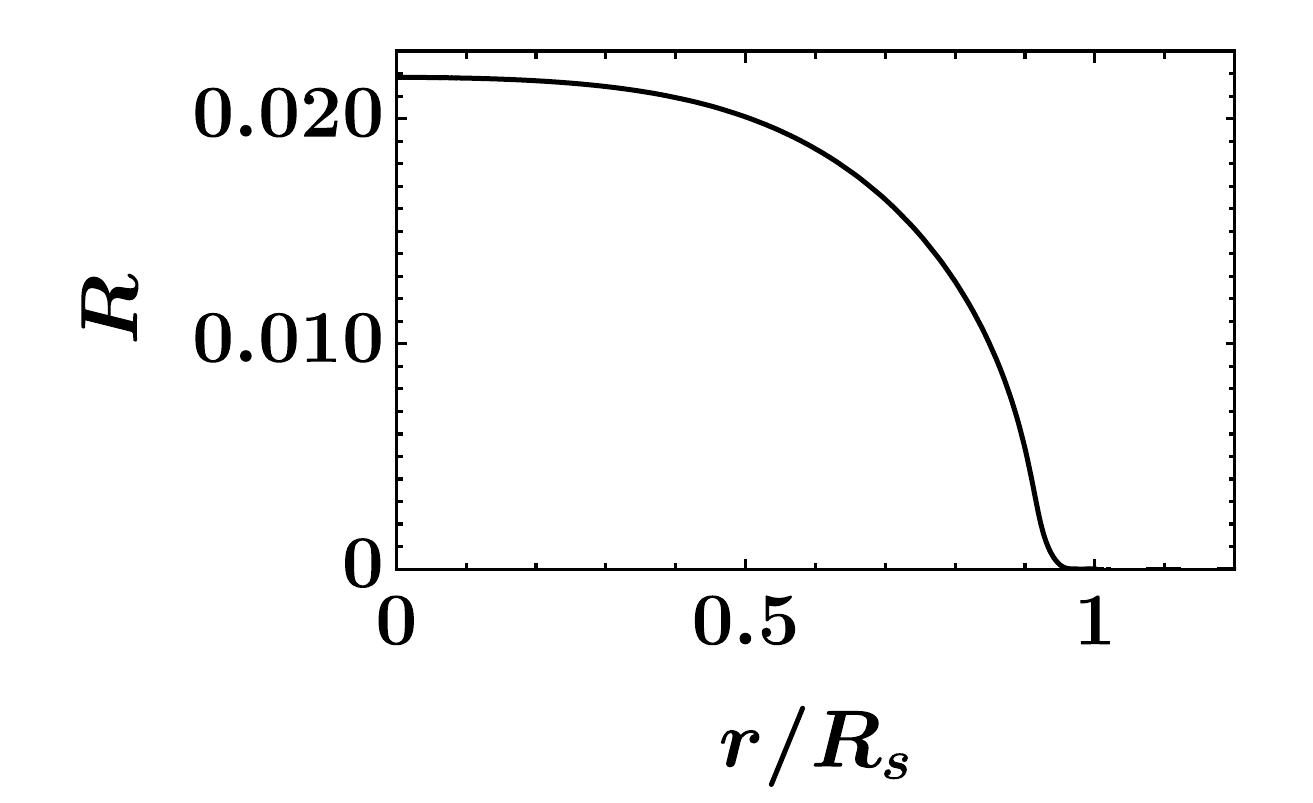}%
}\hfill
\subfloat[]{%
  \includegraphics[width=.49\linewidth]{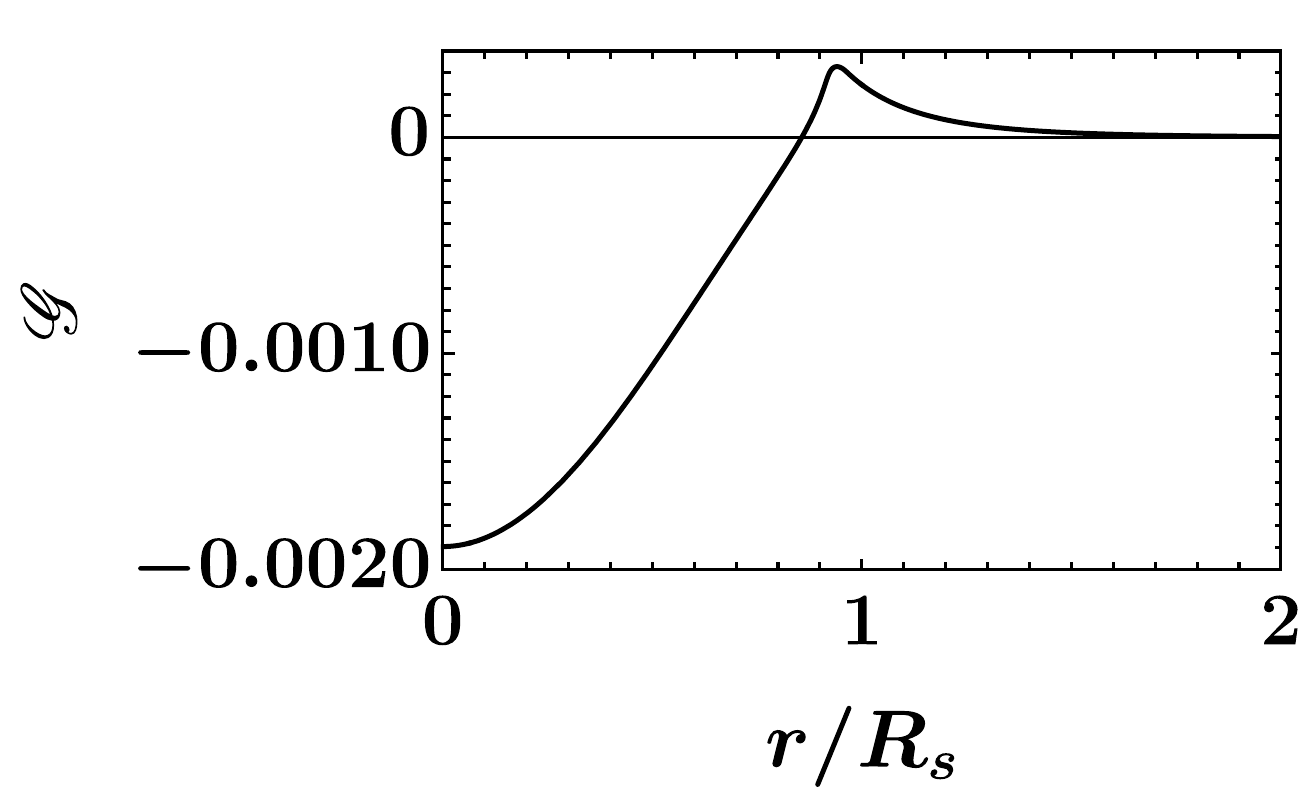}%
}
\caption{Ricci scalar and Gauss-Bonnet invariant for a light star ($M=1.12~M_\odot$, SLy equation of state). The radial coordinate is rescaled by the radius of the star $R_\mathrm{s}=11.7$~km. The left panel shows that the Ricci scalar is non-negative everywhere; correspondingly, only $\beta<0$ can lead to an instability. On the other hand, the Gauss-Bonnet invariant, shown in the right panel, is negative in the core of the star and positive towards its surface, leading to instabilities both for $\alpha<0$ and $\alpha>0$.}
\label{fig:RGB112}
\end{figure}

To understand better the shape of the plot, especially along the $\beta$ and $\alpha$ axes, we plot in Fig.~\ref{fig:RGB112} the Ricci scalar and the Gauss-Bonnet invariant.
The Ricci scalar is always positive on this background. This is due to the fact that we consider a relatively light neutron star, and due to the following relation:
\begin{equation}
R=\kappa(\epsilon-3P).
\label{eq:ricci}
\end{equation}
When the medium is not too dense, $\epsilon\gg P$ and $R>0$. We will see how this changes for a very dense star in sec.~\ref{sec:star}. As a consequence, only negative $\beta$ can generate a negative effective mass. On the other hand, as shown in Fig.~\ref{fig:RGB112}, the Gauss-Bonnet invariant is  positive in some regions and negative in others. This is enough to trigger an instability when $\alpha$ becomes very negative or very positive, which is indeed what is observed in Fig.~\ref{fig:alphabeta}.

We notice, as expected, that the point $(0,0)$ is always inside the stable region. The tachyonic instability does not appear right away when $\beta<0$ or $\alpha\neq0$, due to the curvature of spacetime. 

\subsection{Effect of the bare scalar mass}
\label{sec:mphi}

\begin{figure}
\centering
\includegraphics[width=0.5\textwidth]{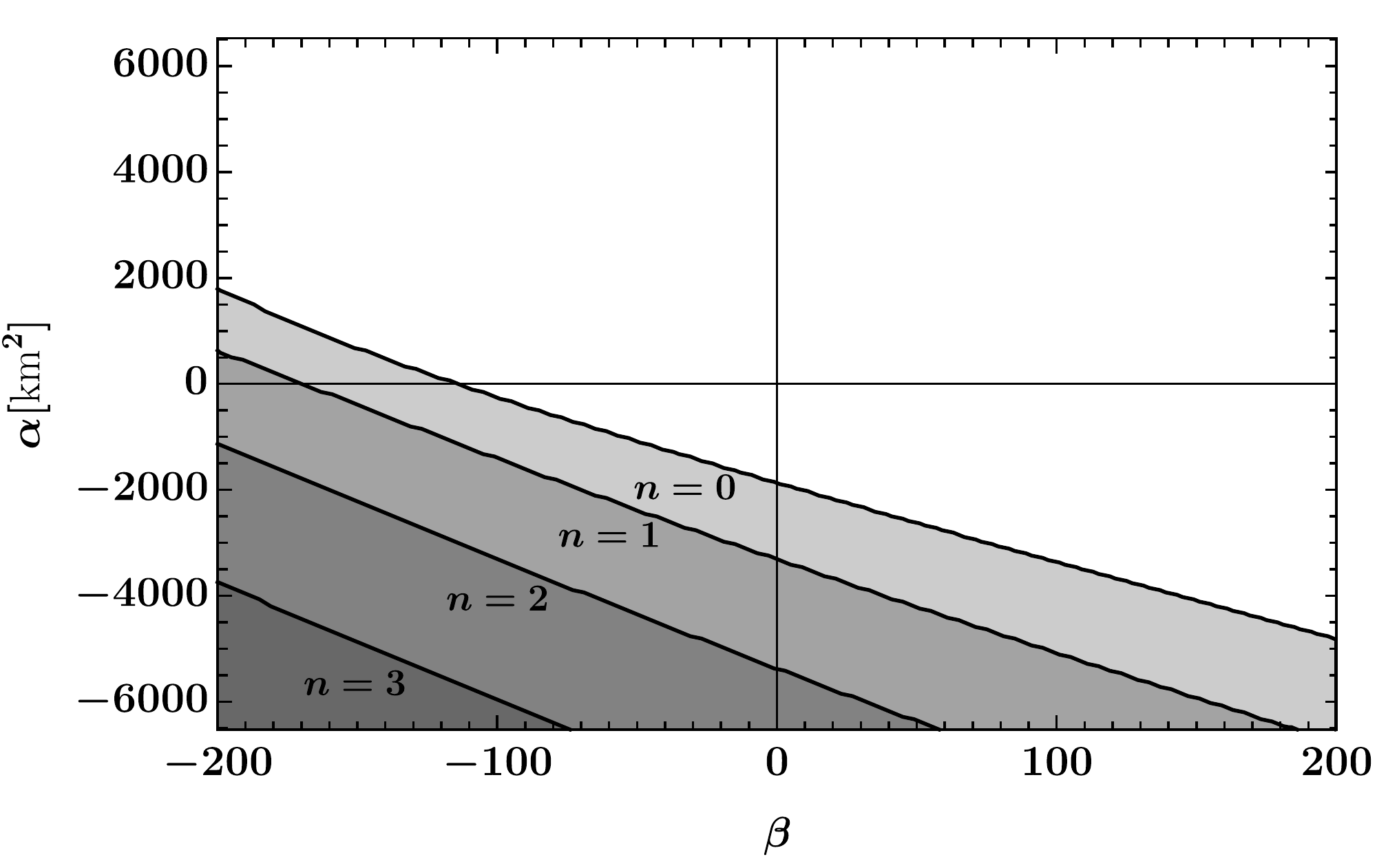}
\caption{Stable and unstable regions in the $(\beta,\alpha)$ space for a bare mass of $1.33\times10^{-10}$~eV. As expected, the stable region is enlarged with respect to Fig.~\ref{fig:alphabeta}. Note that the range of the plot for $\beta$ and $\alpha$ is the same as in Fig.~\ref{fig:alphabeta} to facilitate comparison.}
\label{fig:mphi}
\end{figure}

We now consider how the presence of a bare mass affects the results of the previous section. Fig.~\ref{fig:mphi} shows the region of stability in the $(\beta,\alpha)$ plane when $m_\phi=1$ in the system of units that we used, i.e., scalar particles have a mass of $1.33\times10^{-10}$~eV. The range of parameters in Fig.~\ref{fig:mphi} is the same as in Fig.~\ref{fig:alphabeta} in order to allow comparison. When one zooms out, Fig.~\ref{fig:mphi} looks very similar to Fig.~\ref{fig:alphabeta}  (i.e., the same instability pattern remains valid, but it appears for higher values of the parameters).
As can be seen from comparing Figs.~\ref{fig:alphabeta} and ~\ref{fig:mphi}, the stable region is widened in all directions. As expected, the presence of a bare mass stabilizes the solution (similarly, if the bare mass is tachyonic, the stability region shrinks). Therefore, even a very light bare mass for the scalar field is able to shield neutron stars from scalarization. This was already noted, e.g., in~\cite{Ramazanoglu:2016kul}, where the effect of a bare mass in a cosmological setup is also discussed.

\section{Changing the properties of the star}
\label{sec:star}

Let us now examine how changing the background affects the stability. We first consider a more massive star, and then a different equation of state.

\subsection{Mass of the star}

\begin{figure}
\centering
\includegraphics[width=0.5\textwidth]{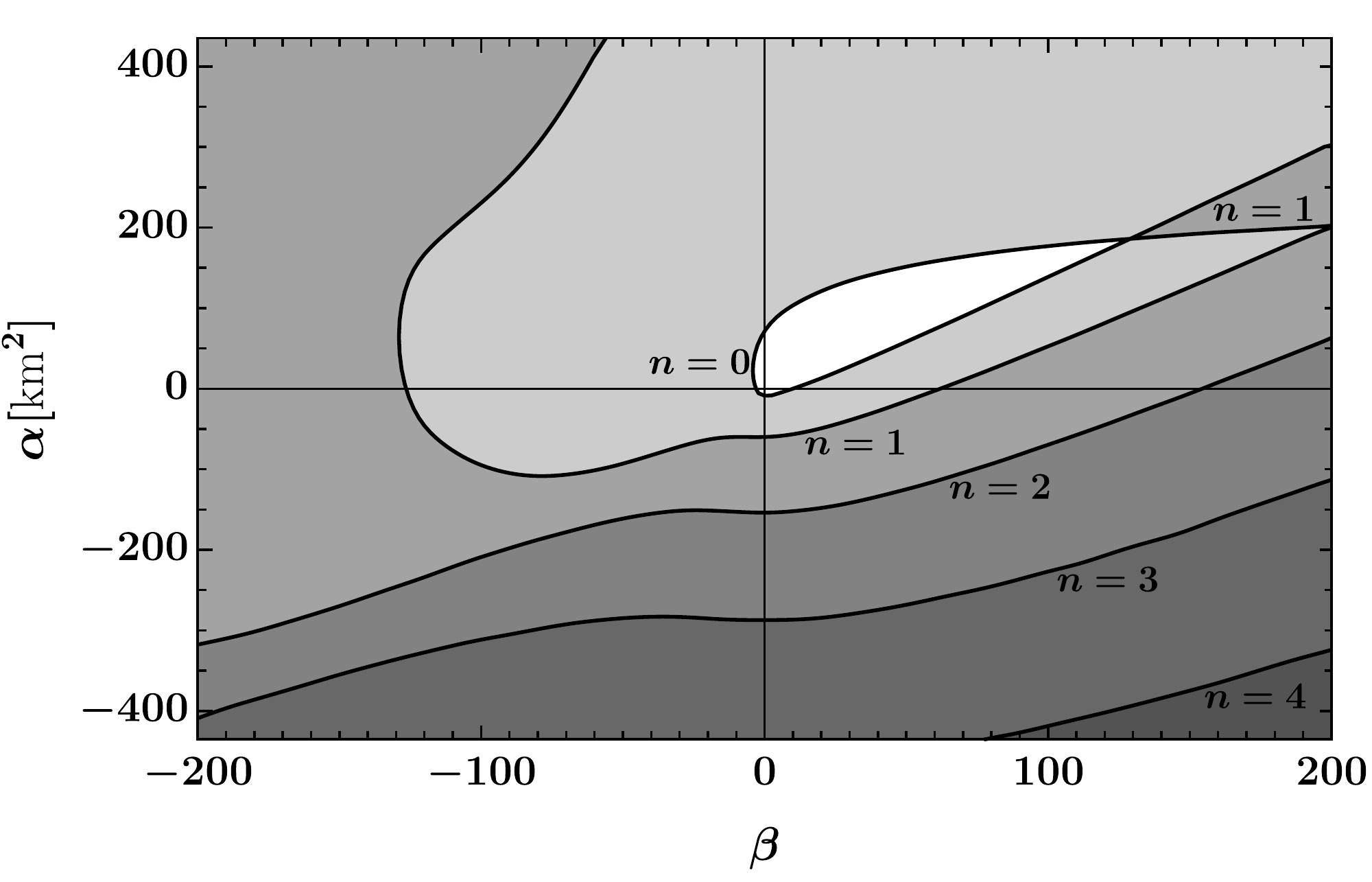}
\caption{Stable and unstable regions in the $(\beta,\alpha)$ space for a heavy neutron star ($M=2.04~M_\odot$, SLy equation of state). The vertical range for $\alpha$ is reduced with respect to Fig.~\ref{fig:alphabeta} for readability. The wedge of stability in Fig.~\ref{fig:alphabeta} has narrowed down to an island around $(\beta,\alpha)=(0,0)$.}
\label{fig:heavystar}
\end{figure}

\begin{figure}
\subfloat[]{%
  \includegraphics[width=0.49\linewidth]{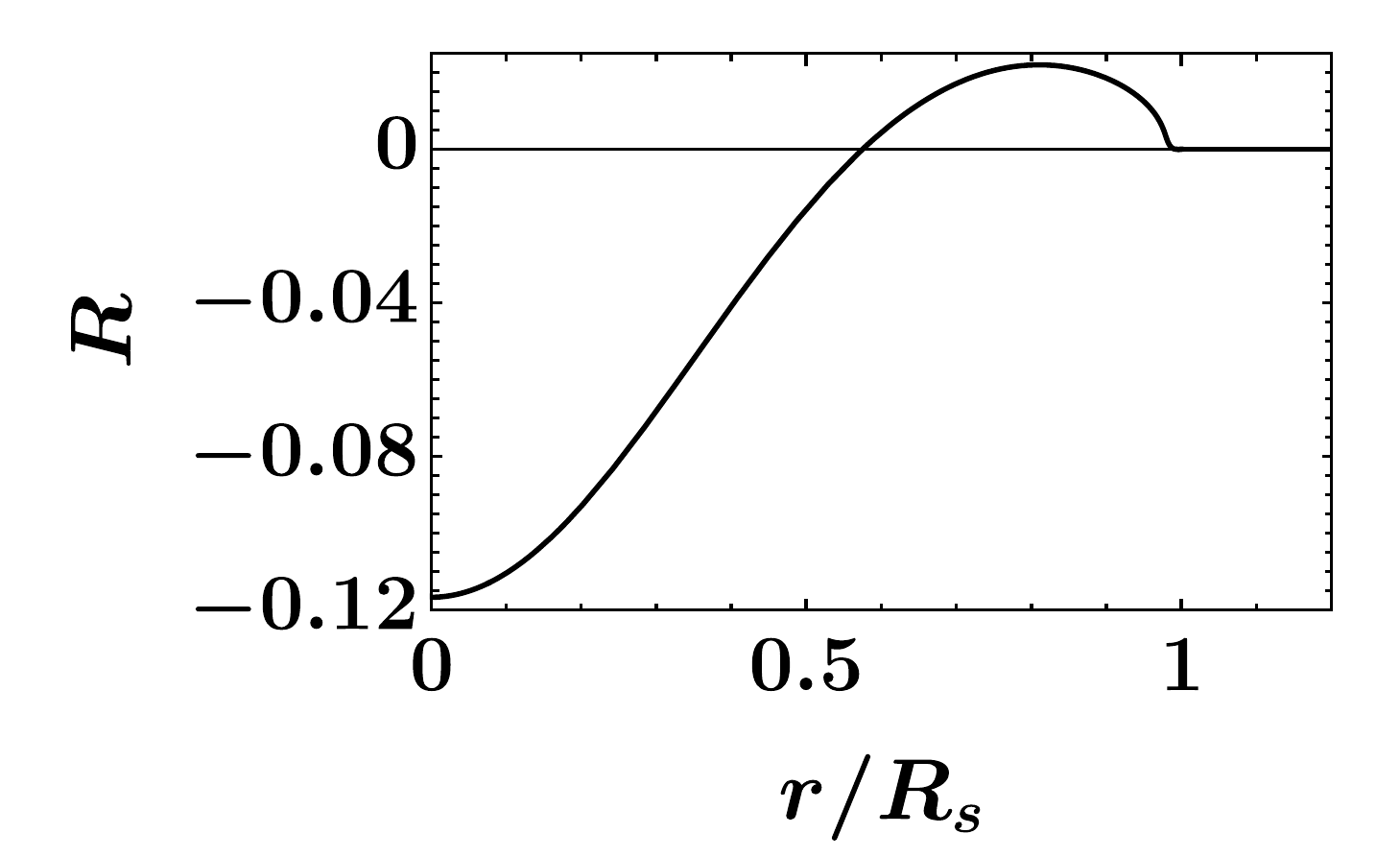}%
}\hfill
\subfloat[]{%
  \includegraphics[width=.49\linewidth]{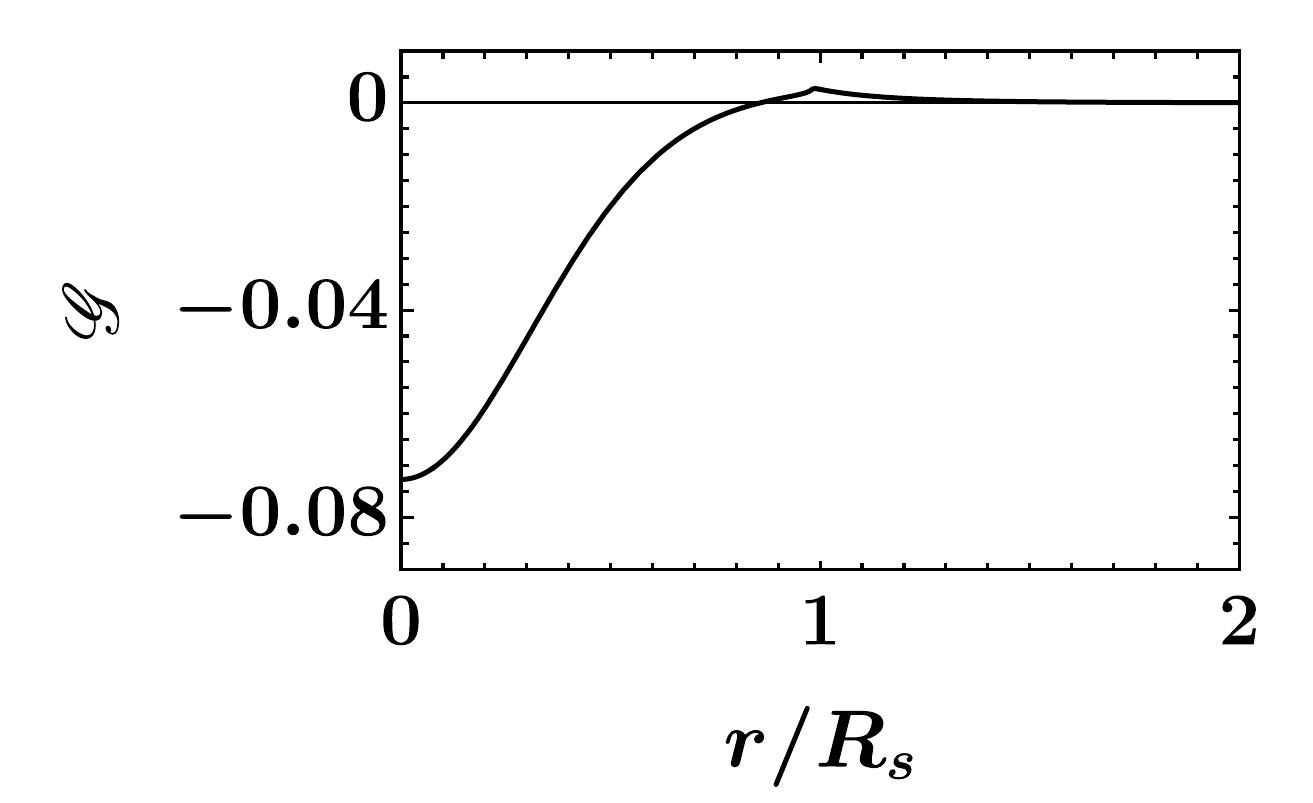}%
}
\caption{Ricci scalar and Gauss-Bonnet invariant for a heavy star ($M=2.04~M_\odot$, SLy equation of state). Now, the left panel shows that $R$ becomes negative inside the core of the star, allowing instabilities for both signs of $\beta$. The Gauss-Bonnet invariant (right panel) still behaves as in the case of a light star, Fig.~\ref{fig:RGB112}.}
\label{fig:RGB204}
\end{figure}

We first increase the mass of the neutron star (and thus its compactness at the same time). We choose a central density of $\rho_\mathrm{c}=3.4\times10^{18}$~kg/m$^3$, which corresponds to a mass of $M=2.04~M_\odot$. This is the heaviest spherically symmetric star we can produce with the SLy equation of state; beyond this mass, the solutions become unstable already within GR. The results are displayed in Fig.~\ref{fig:heavystar}.
Since the curvature of the background increased with respect to Fig.~\ref{fig:alphabeta}, it is not surprising that the stable region shrunk. A more specific feature is that very positive values of $\beta$ now also lead to an instability. This is due to the fact that the Ricci scalar now becomes negative towards the center of the star, as can be seen in Fig.~\ref{fig:RGB204}.
In terms of eq.~\eqref{eq:ricci}, the energy density is no longer dominant with respect to the pressure in the extremely pressurized core of the neutron star, allowing $R<0$. This effect was already noted in \cite{Mendes:2014ufa}, and further studied in \cite{Palenzuela:2015ima,Mendes:2016fby}.

\subsection{Equation of state}

\begin{figure}
\centering
\includegraphics[width=0.5\textwidth]{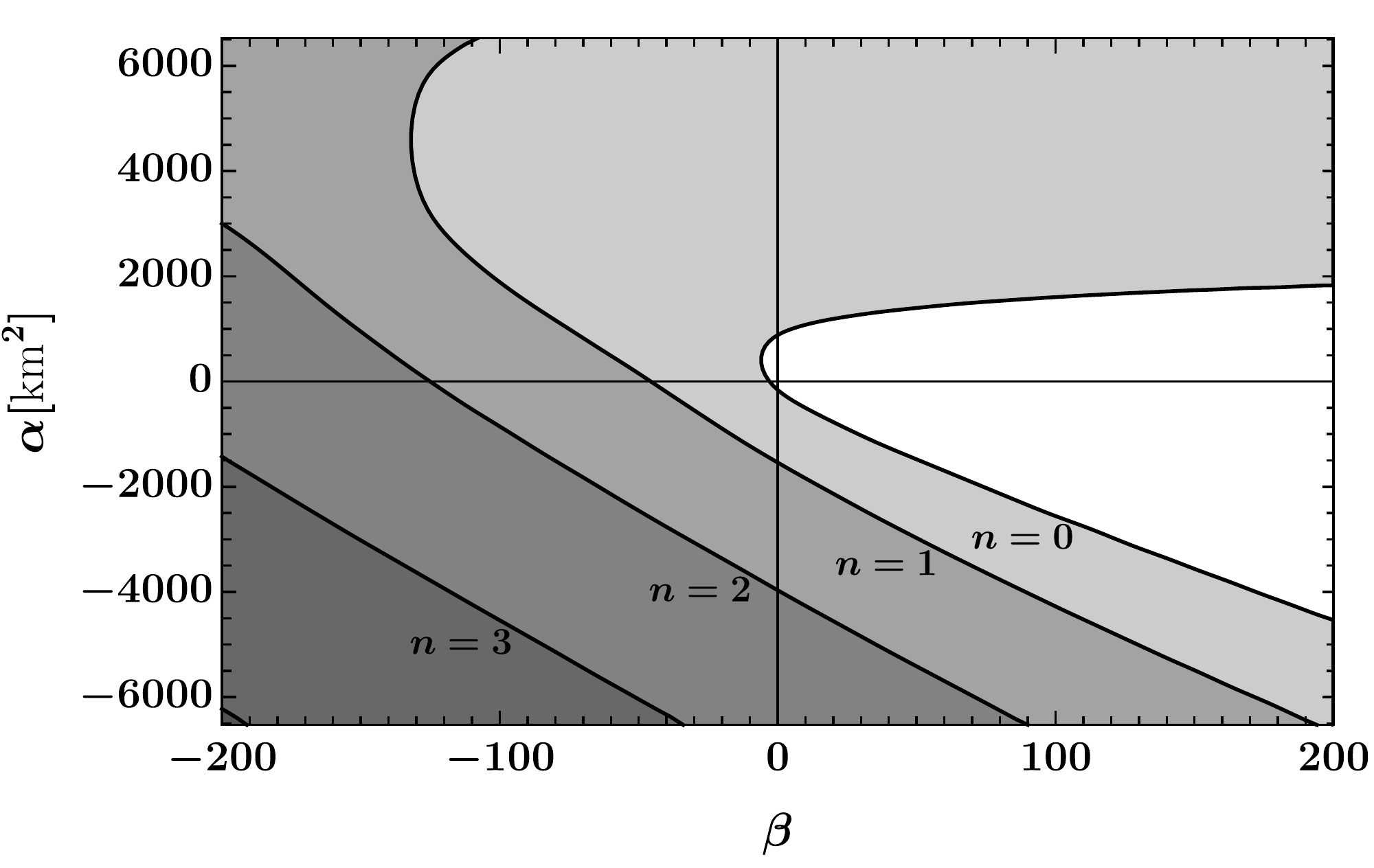}
\caption{Stable and unstable regions in the $(\beta,\alpha)$ space for the MPA1 equation of state ($M=1.12~M_\odot$). The range for $\beta$ and $\alpha$ is the same as in Fig.~\ref{fig:alphabeta}. We note that the choice of equation of state does not affect significantly the results.}
\label{fig:MPA1}
\end{figure}

We now consider a different stellar model, the MPA1 equation of state~\cite{Gungor:2011vq}. In order to compare with previous results, we keep the same mass as in Sec.~\ref{sec:effmass}, $M=1.12~M_\odot$. This corresponds to a central density of $\rho_\mathrm{c}=6.3\times10^{17}$~kg/m$^3$. The stability region is shown in Fig.~\ref{fig:MPA1}.
By comparing Figs.~\ref{fig:alphabeta} and \ref{fig:MPA1}, it is clear that the different choice of equation of state does not affect significantly the position of the stable and unstable regions. 

Although the equation of state influences only mildly stars of similar mass, it can have indirect effects. Indeed, a softer equation of state will lead to a smaller pressure for a given energy density; thus, for a soft equation of state, the Ricci scalar \eqref{eq:ricci} could remain positive in all configurations, discarding configurations like Fig.~\ref{fig:heavystar}. Similarly, the equation of state can affect the range allowed for the parameter $\gamma$, as we will see in more detail in Sec.~\ref{sec:a41}.


\section{Changing the effective metric}
\label{sec:a41}
We now return to the parameter $\gamma$ and examine its role. As mentioned above, the terms controlled by $\gamma$ cannot  generate a tachyonic instability in the absence of the other couplings controlled by $\beta$, $\alpha$ or $m_\phi$. Nonetheless, choosing $\gamma$ beyond a certain range leads to loss of hyperbolicity in the scalar field equation. Also the potential $V_\mathrm{eff}$ does depend on $\gamma$, as can be seen from eq.~\eqref{eq:Veff}. We analyze these aspects below, before studying numerically the combined effect of $\gamma$ and the other parameters.

\subsection{Hyperbolicity}

The parameter $\gamma$ brings an additional contribution to the effective metric experienced by scalar perturbations, eq.~\eqref{effMetr}. If this contribution exceeds a certain threshold and becomes dominant, the effective metric becomes elliptic, rendering the background unstable.  We emphasize that this instability is distinct from the usual tachyonic instability associated with scalarization. Depending on the circumstances (see also below), it is either a gradient or a ghost instability. It is possible that this instability can be quenched nonlinearly, as is the case for conventional scalarization. Indeed, scalarization through a ghost instability has already been proposed in \cite{Ramazanoglu:2017yun}. Since here we are not including terms that are nonlinear in the scalar in our analysis, we cannot follow the development and potential quenching.

One can therefore view the analysis that follows in two different ways. Taking the conservative viewpoint, one may restrict to the well-controlled tachyonic scalarization. In this framework, our results allow to set bounds on the parameter $\gamma$. In a more open-minded perspective, which certainly deserves further investigation in the future, we are setting  bounds beyond which ghost or gradient scalarization can be triggered.

\begin{figure}
\centering
\includegraphics[width=0.5\textwidth]{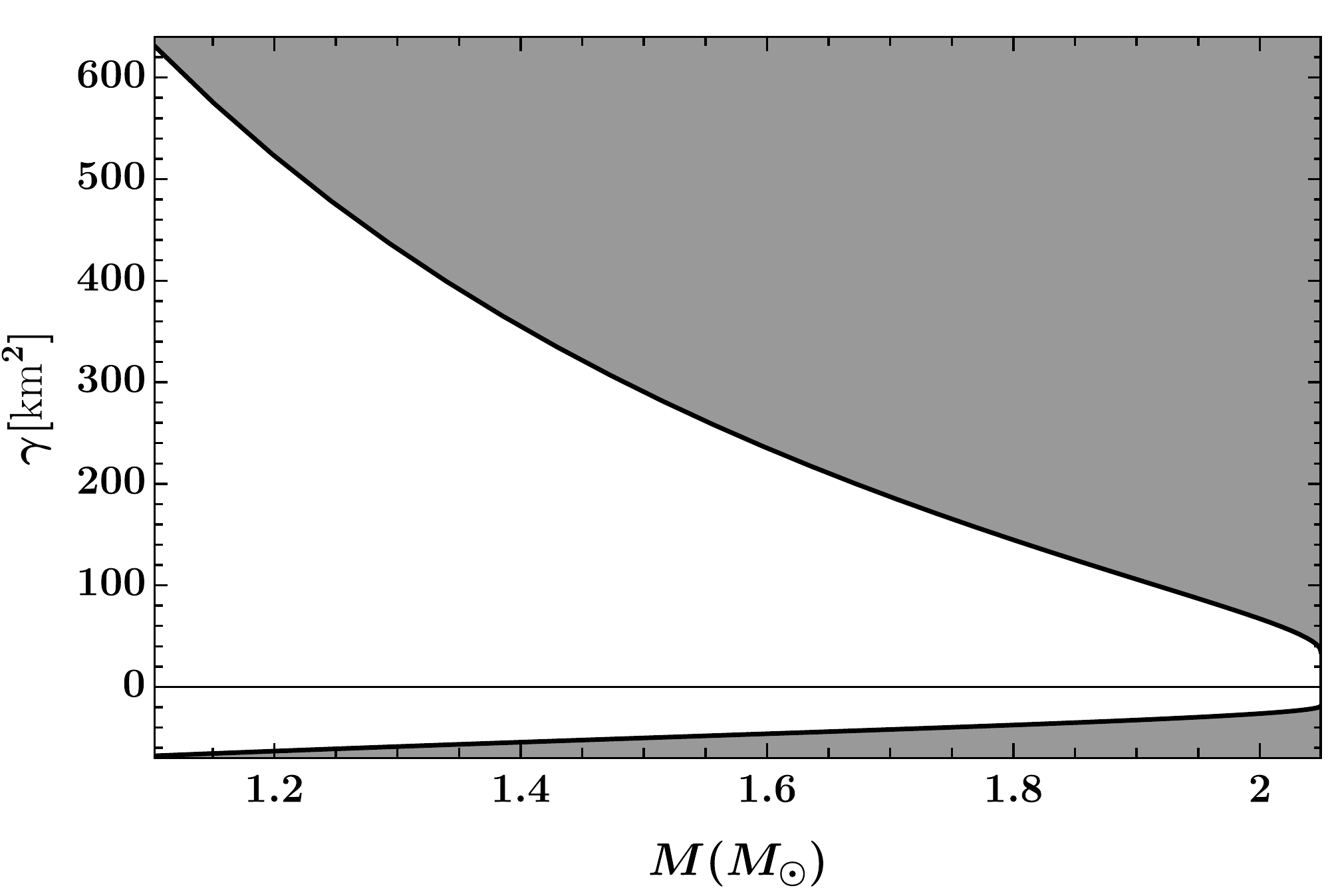}
\caption{Hyperbolicity of the effective metric in the $(M,\gamma)$ plane (SLy equation of state). The effective metric is hyperbolic only within the white region. The mass ranges from the putative lowest neutron star mass \cite{Martinez:2015mya,Suwa:2018uni} to the maximal mass achieved with the SLy equation of state. The range of hyperbolicity narrows down when increasing the mass of the star.}
\label{fig:Ma41}
\end{figure}

Given that we are working on a GR background, we can make use of Einstein equations. The inverse of $\tilde g^{\mu\nu}$ for a perfect fluid is then 
\begin{equation}
\tilde g^{-1}_{\mu\nu} = \dfrac{1}{1-\kappa \gamma P} \left(g_{\mu\nu}-\kappa \gamma\dfrac{\epsilon+P}{1+\kappa \gamma\epsilon}\,u_\mu u_\nu\right),
\end{equation}
where $u_\mu$ is the 4-velocity of the fluid. Over the spherically symmetric background that we study, the determinant of the effective metric reads
\begin{equation}
\tilde g = g \dfrac{1}{(1-\kappa \gamma P)^3(1+\kappa \gamma\epsilon)}\,.
\end{equation}
On a given background, the pressure and energy density are maximal at the center of the star, where we label their value as $P_\mathrm{c}$ and $\epsilon_\mathrm{c}$. The determinant of the physical metric $g$ is always negative. Thus, the effective metric loses hyperbolicity either when $\kappa \gamma$ becomes larger than $1/P_\mathrm{c}$ or when $\kappa \gamma$ becomes more negative than $-1/\epsilon_\mathrm{c}$. To summarize, hyperbolicity of the effective metric requires that
\begin{equation}
-\dfrac{1}{\kappa \epsilon_\mathrm{c}} <\gamma<\dfrac{1}{\kappa P_\mathrm{c}}.
\label{eq:bounda41}
\end{equation}
Reference~\cite{Minamitsuji:2016hkk} already noted the existence of the upper bound in a similar context. In the numerical analysis below, we will take the conservative approach and restrict to this range. For a given equation of state, the bounds in eq.~\eqref{eq:bounda41} can be reformulated in terms of the mass of the star, $M$. This is shown in Fig.~\ref{fig:Ma41}.
The range in which $\tilde g_{\mu\nu}$ is hyperbolic closes up around $\gamma=0$ when increasing the mass. The limits presented in Fig.~\ref{fig:Ma41} depend on the equation of state, but only mildly. In the framework of tachyonic scalarization, we can use these limits to put an absolute bound on $\gamma$. For the SLy equation of state,
\begin{equation}
-18.7~\mathrm{km}^2 <\gamma<34.9~\mathrm{km}^2.
\label{eq:absbounda41}
\end{equation}
These bounds should only be taken as order of magnitude estimates; they have been established in the decoupling limit and rely on a specific equation of state. However, a more detailed study would allow to put very precise constraints on $\gamma$. We are not aware of previously established bounds on this parameter (except \cite{Minamitsuji:2016hkk}). Note that the analysis presented in this paragraph is entirely independent on the other parameters, which do not play a role in the effective metric.

\subsection{Integral of the effective potential}

Before moving to the numerical results, we point out an important difference between black holes and stars. In the case of black holes, the scalar field equation can also be brought to the form \eqref{scalarEqPot2}. In both cases (black hole and neutron star), we impose that $\mathrm{d}\hat\sigma/\mathrm{d}r_\ast$ vanishes at large $r_\ast$ so that $\int \phi^2$ is finite and the perturbation initially contains a finite amount of energy. The difference between neutron stars and black holes appears on the other side of the $r_\ast$ range. In the case of neutron star, $r_\ast$ goes down to 0, where $\phi=K\hat\sigma\sim\phi_0/r_\ast$ for some constant $\phi_0$, unless $\hat\sigma(0)=0$. We do not want $\phi$ to diverge at the center of the star, so we impose $\hat\sigma(0)=0$. In the black hole case on the other hand, $r_\ast$ goes down to $-\infty$ and nothing particular happens at $r_\ast=0$. The condition for the perturbation to be physical is then that $\mathrm{d}\hat\sigma/\mathrm{d}r_\ast$ vanishes for $r_\ast\to-\infty$, similarly to what happens at large $r_\ast$.

It is then possible to establish an exact sufficient condition for the presence of an instability in the black hole case. Indeed, the function $\hat\sigma$ then respects the hypotheses of the theorem established in ref.~\cite{doi:10.1119/1.17935}: if $\int_{-\infty}^{+\infty} V_\mathrm{eff}(r_\ast)\mathrm{d}r_\ast<0$, then $V_\mathrm{eff}$ admits at least one bound state. 

However, in the case of a star, due to the different boundary conditions, $\int_{0}^{+\infty} V_\mathrm{eff}(r_\ast)\mathrm{d}r_\ast$ can become negative ---~even infinitely negative~--- while $V_\mathrm{eff}$ does not possess any bound state. This is equivalent to the 3D-spherically symmetric quantum well; consider such a well with depth $-V_0$ and width $a$. It admits a bound state only for $V_0a^2>N$ ($N$ is some constant that depends on the mass of the quantum particles), while the integral of the potential is $-V_0 a<0$. Choosing the scaling $V_0=N/2/a^2$, the integral of the potential is then becoming infinitely negative for $a\to0$, while no bound state exists.

Hereinafter, $\int V_\mathrm{eff}(r_\ast)\mathrm{d}r_\ast$ indeed becomes infinitely negative in the limit where $\gamma$ approaches one of the bounds of eq.~\eqref{eq:bounda41}. However, this does not necessarily mean that infinitely many bound states develop close to these boundaries. This seems to be true close to the upper bound, but not close to the lower one, as we will see in Fig.~\ref{fig:alphaa41}.

\subsection{Numerical analysis}
\label{sec:numa41}

\begin{figure}
\centering
\includegraphics[width=0.5\textwidth]{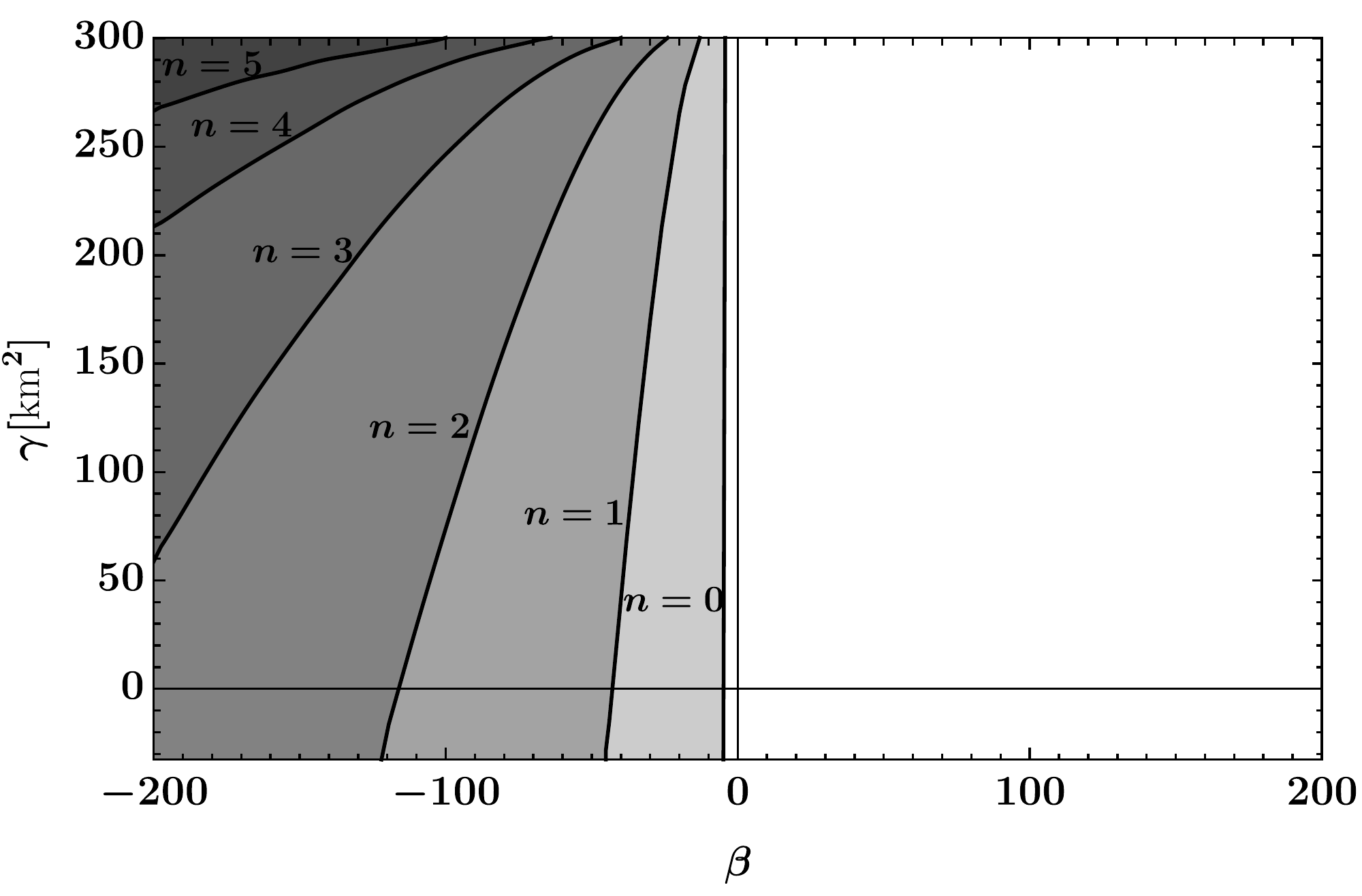}
\caption{Stable and unstable regions in the $(\beta,\gamma)$ space ($M=1.12~M_\odot$, SLy equation of state). The region of stability is rather unaffected by a change in $\gamma$. The bound states pile up close to the upper bound for $\gamma$.}
\label{fig:alphaa41}
\end{figure}

\begin{figure}
\centering
\includegraphics[width=0.5\textwidth]{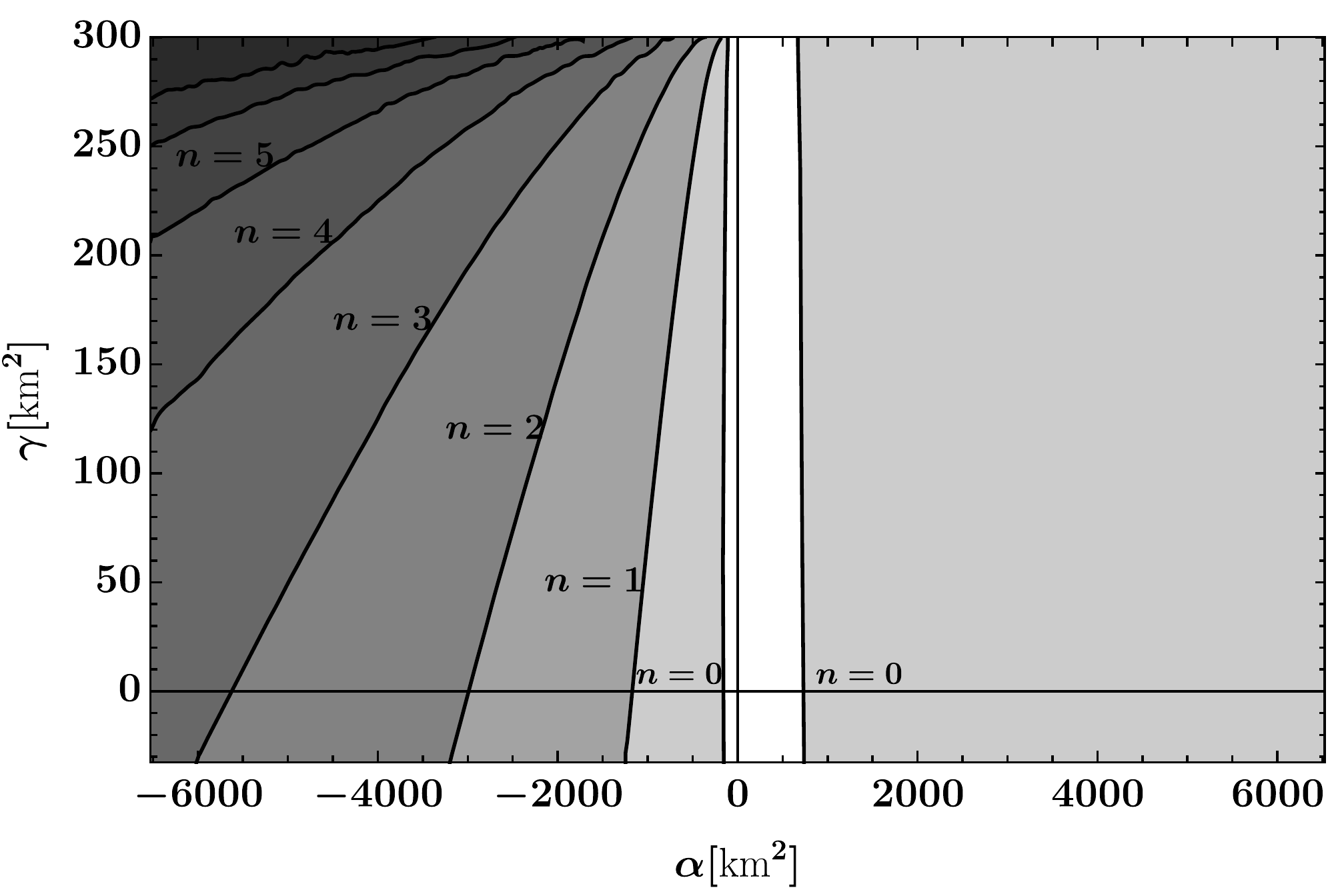}
\caption{Stable and unstable regions in the $(\alpha,\gamma)$ space ($M=1.12~M_\odot$, SLy equation of state). The behavior is very similar to what we obtained for $\beta$ in Fig.~\ref{fig:alphaa41}.}
\label{fig:betaa41}
\end{figure}

We now vary $\gamma$ systematically in the range allowed by eq.~\eqref{eq:bounda41}, and $\beta$ and $\alpha$ in similar ranges as in the previous figures. The results are displayed in figs.~\ref{fig:alphaa41} and \ref{fig:betaa41} respectively. 
When $\gamma$ vanishes (cut along the horizontal axis in figs.~\ref{fig:alphaa41} and \ref{fig:betaa41}), we retrieve the same stability ranges as in the left and bottom panels of Fig.~\ref{fig:alphabeta} ($\beta>-5.42$ and $722~\mathrm{km}^2>\alpha>-169~\mathrm{km}^2$). It is also natural that the lines $\beta=0$ and $\alpha=0$ (vertical axes in figs.~\ref{fig:alphaa41} and \ref{fig:betaa41}) correspond to stable configurations, because $\gamma$ cannot create a tachyonic effective mass on its own. The boundary of the stable region is rather insensitive to the parameter $\gamma$; it does evolve slightly with the value of $\gamma$, but this can be seen only when zooming around smaller values of $\beta$ and $\alpha$ with respect to figs.~\ref{fig:alphaa41} and \ref{fig:betaa41}. Close to both bounds of the $\gamma$ range, $\int V_\mathrm{eff}(r_\ast)\mathrm{d}r_\ast$ diverges, but as explained in the previous section, this does not necessarily mean that infinitely many bound states should appear (or even that one bound state exists). Indeed, nothing particular happens when approaching the lower bound, while unstable modes pile up when approaching the upper bound; both scenarios are allowed.

\section{Conclusion}
\label{sec:conc}

We have investigated exhaustively the effect of all the terms that can play a role in the onset of spontaneous scalarization, within the context of Horndeski theory. Our analysis has identified the role of each term but has also revealed their combined effects. When taking each terms separately, our results agree with previous results regarding scalarization thresholds. More generally and when all terms are present, we have found that  a very small bare mass suffices to stabilize GR solutions, and that the scalarization thresholds are only mildly sensitive to the choice of equation of state.

\begin{figure}
\includegraphics[width=0.45\textwidth]{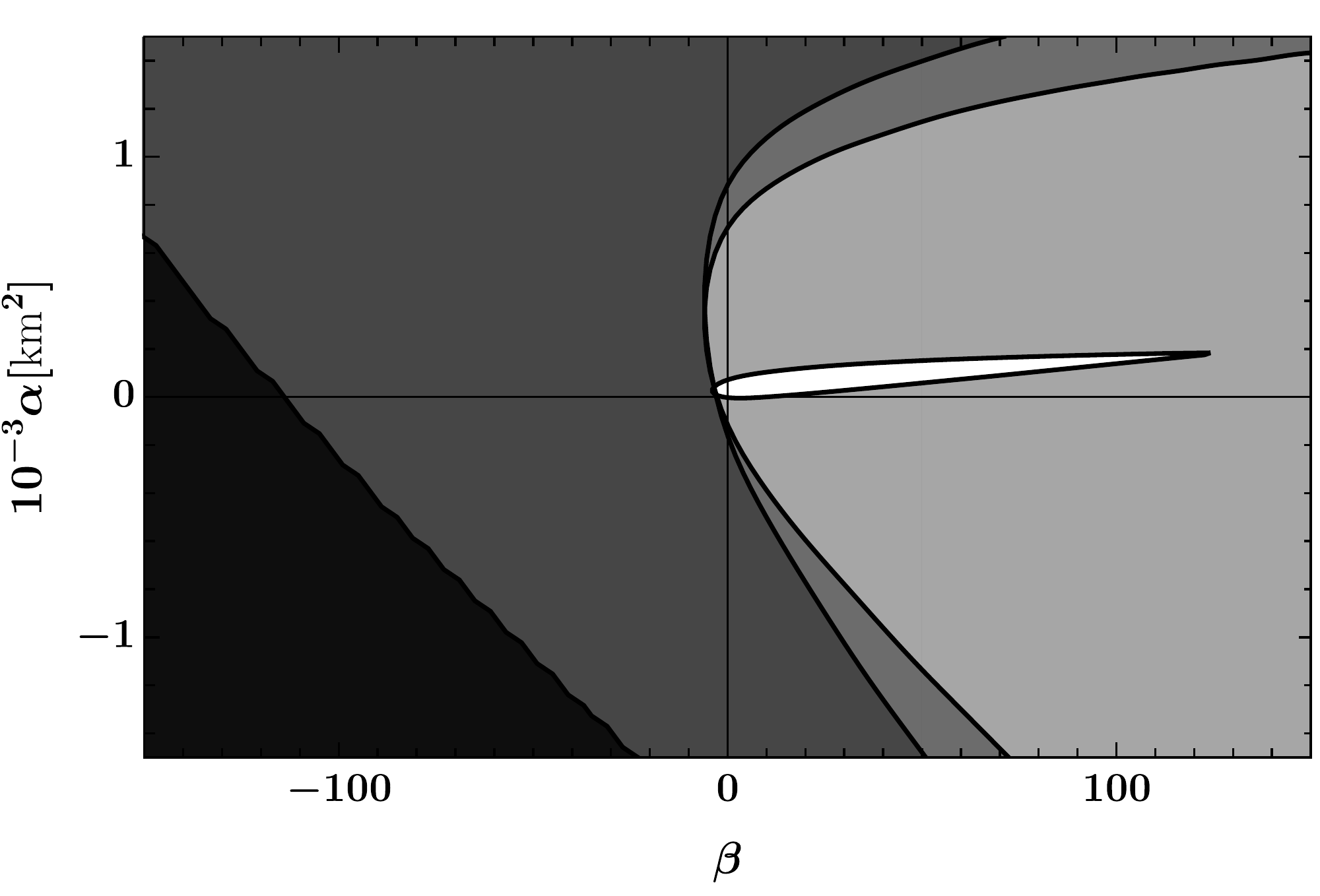}
\caption{Summary of the stability ($n=0$) contours in the $(\beta,\alpha)$ space
for comparison. The white region corresponds to the region of stability for a heavy star, $M=2.04~M_\odot$, using the SLy equation of state. The light grey region (together with the white region) is the stability region for $M=1.12~M_\odot$, still with the SLy equation of state. The intermediate grey region (together with the previous paler regions) corresponds to a mass $M=1.12~M_\odot$ and the MPA1 equation of state. The dark grey region (again, with paler regions) corresponds to a mass $M=1.12~M_\odot$ and the SLy equation of state, together with a bare scalar mass of $1.33\times10^{-10}$~eV. Finally, the black region is unstable for all the previous models.}
\label{fig:thresholds}
\end{figure}

Our analysis allowed us to explore, for the first time, the multi-dimensional parameter space and provide scalarization thresholds that depend on more than one coupling. In Fig.~\ref{fig:thresholds} we presented in a single plot a summary of most of the stability contours presented in this paper. We expect that our multi-parameter analysis will be particularly useful in building viable scalarization models. Observational constraints on these models will come from different compact objects, such as black holes and neutron stars, and also from low curvature observations. The Ricci scalar and the Gauss-Bonnet invariant, which appear in the terms that contribute to the onset of scalarization, scale differently with curvature and are not sign-definite. Hence, by choosing the sign and magnitude of different couplings constants in the minimal action of eq.~\eqref{eq:ActionCaseI} one can construct models in which only certain types of compact objects are scalarized and also control the behaviour of the model at lower curvatures. A characteristic example in this direction has recently been discussed in Ref.~\cite{Antoniou:2020nax}: a suitable choice of parameters yielded a black hole scalarization model with GR as a cosmological attractor. 

One of the striking features revealed by our analysis is 
the role of the effective metric in which scalar perturbations propagate. It is controlled by a single coupling constant, $\gamma$. There exists a threshold beyond which the effective metric loses hyperbolicity. In the framework of tachyonic scalarization, we interpret this threshold as an absolute bound on the parameter $\gamma$. It is restricted to a rather narrow range (roughly, it should remain small with respect to the characteristic length of curvature). Therefore, it has a very limited effect on the threshold of tachyonic scalarization. On the other hand, the loss of hyperbolicity can be seen as an alternative instability that could lead to scalarization, in line with what was proposed in Ref.~\cite{Ramazanoglu:2017yun} in a more restricted setup. It would be interesting to understand if such an instability can indeed be controlled and give rise to a sensible scalarization process. 

More generally, the next natural step is to take into account non-linearities in the scalar field, which are excluded from the minimal action \eqref{eq:ActionCaseI} by definition. They determine the end state of any scalarization instability and control the deviations of scalarized compact objects from their GR counterparts. A detailed study would encompass the existence of scalarized solutions, their stability,  their transition from the GR branch to the scalarized branch, and the associated observational signatures and constraints. We leave these for future work. 

\begin{acknowledgements}
We thank Eugeny Babichev for useful discussions. AL and TPS would like to acknowledge partial support from STFC grant ST/P000703/1. We would also like to acknowledge networking support by the COST Action GWverse CA16104.
\end{acknowledgements}
\appendix

\section{Background equations}
\label{sec:TOV}

Matter is described as a perfect fluid with stress-energy tensor
\begin{equation}
T_{\mu\nu}=(\epsilon+P)u_\mu u_\nu+P g_{\mu\nu},
\end{equation}
where $\epsilon$ is the energy density of the fluid, $P$ its pressure and $u_\mu$ its 4-velocity. The system of coordinates \eqref{eq:gansatz} is chosen so that the fluid is at rest. Therefore,
\begin{equation}
u_\mu=(-c\sqrt{h},0,0,0).
\end{equation}
The local mass density is defined as $\rho=\epsilon/c^2$. In this setup, Einstein's field equations take the form
\begin{align}
0&=(r f)'- 1 +\kappa \epsilon r^2,
\\
0&=\dfrac{f}{h}(rh)'- 1 -\kappa P r^2 .
\end{align}
Additionally, the conservation equation $\nabla_\mu T^{\mu\nu}=0$ can be put in the form
\begin{equation}
0=-\dfrac{1}{2 r f}[(-1 + f - \kappa P r^2) (P + \epsilon)] + P'.
\end{equation}
Together with an equation of state $P(\epsilon)$, these equations allow to solve for the background geometry and the matter distribution. Note that the equation of states we used are 23 parameters fits of the actual equations of state~\cite{Gungor:2011vq}.

\section{The effective potential}
\label{App:Veff}

In terms of the background functions $f$ and $h$ and the parameters $m_\phi$, $\beta$, $\alpha$ and $\gamma$, the effective potential presented in sec.~\ref{sec:setup} reads
\begin{widetext}
\begin{equation}
\begin{split}
V_\text{eff}(r) &= \bigg\{ h \{4 \gamma  f^2 h'^2 \{[-2 \beta  r^2+r^2-2 \gamma  f' r+16 \alpha +2 \gamma -8 (2 \alpha +\gamma ) f] h'-2 r \gamma  f h''\} r^3\\
&\quad +f h \{\{-16 \gamma  (4 \alpha +5 \gamma ) f^2+4 \{8 \alpha  (r^2+4 \gamma )+\gamma [(6 \beta +1) r^2+14 \gamma ]+4 r \gamma  [(12 \alpha +5 \gamma ) f'+r \gamma  f'']\} f\\
&\quad +[(4 \beta -3) r^2-32 \alpha -6 \gamma ] (r^2+2 \gamma )+4 r \gamma  f' [(2 \beta +1) r^2+\gamma  f' r+2 (\gamma -8 \alpha )]\} h'^2\\
&\quad +8 r \gamma  f [(2 \beta  r^2+4 \gamma  f' r-16 \alpha +16 \alpha  f+5 \gamma  f) h''+2 r \gamma  f h^{(3)}] h'-4 r^2 \gamma ^2 f^2 h''^2\} r^2\\
&\quad +2 h^2 \{8 \gamma  \{r [4 (2 \alpha +\gamma ) h''+r \gamma  h^{(3)}]-5 \gamma  h'\} f^3+4 \{\gamma  h' [8 \beta  r^2+4 \gamma  f'' r^2+3 (8 \alpha +5 \gamma ) f' r+10 \gamma ]\\
&\quad +r \{-2 [(4 \alpha -\beta  \gamma +2 \gamma ) r^2-2 \gamma ^2 f' r+\gamma  (16 \alpha +5 \gamma )] h''-r \gamma  (r^2+2 \gamma ) h^{(3)}\}\} f^2\\
&\quad +2 r \{(r^2+2 \gamma ) [(1-2 \beta ) r^2-5 \gamma  f' r+2 (8 \alpha +\gamma )] h''-h' \{2 (2 \beta -1) r^3+4 (2 m_\phi^2 r^2+4 \beta -1) \gamma  r\\
&\quad +2 \gamma  (r^2+2 \gamma ) f'' r+f' \{-2 r f' \gamma ^2+5 [(3-2 \beta ) r^2+6 \gamma ] \gamma +8 \alpha  (3 r^2+8 \gamma )\}\}\} f\\
&\quad +r (r^2+2 \gamma ) f' [(1-2 \beta ) r^2-2 \gamma  f' r+2 (8 \alpha +\gamma )] h'\} r+4 h^3 \{-12 \gamma ^2 f^3\\
&\quad +4 \gamma  [2 \beta  r^2+r^2+\gamma  (2 f'+r f'') r+6 \gamma ] f^2+\{-4 \beta  (r^2+4 \gamma ) r^2+\gamma  \{f' [(8 \beta -6) r^2+\gamma  f' r-12 \gamma ]\\
&\quad -2 r (r^2+2 \gamma ) f''\} r-4 \gamma  (2 m_\phi^2 r^4+r^2+3 \gamma )\} f+r (r^2+2 \gamma ) \{-r \gamma  f'^2+2 [(1-2 \beta ) r^2+\gamma ] f'\\
&\quad +4 r (m_\phi^2 r^2+\beta )\}\}\} [r^2+2 \gamma -2 \gamma  (f+r f')]^2-5 f [(r^2+2 \gamma -2 \gamma f) h-2 r \gamma  f h']^2 \{r [r^2+2 \gamma \\
&\quad -2 \gamma  (f+r f')] h'+2 \gamma  h (f'' r^2-2 f+2)\}^2-2 [-r^2-2 \gamma +2 \gamma  (f+r f')] [(-r^2-2 \gamma +2 \gamma  f) h\\
&\quad +2 r \gamma  f h'] \{-4 \gamma ^2 [16 h^3-4 r (r h''-6 h') h^2-3 r^2 h' (r h''-5 h') h+5 r^3 h'^3] f^3\\
&\quad +2 \gamma  \{4 \{3 r^2+\gamma  [4 f'+r (2 f''-r f^{(3)})] r+16 \gamma \} h^3-2 r \{r [2 r^2+\gamma  (r f''-2 f') r+6 \gamma ] h''+h' \{-5 r^2\\
&\quad +\gamma  [r (2 r f^{(3)}-7 f'')-8 f'] r-34 \gamma \}\} h^2+r^2 h' \{h' [9 r^2+2 \gamma  (5 r f''-3 f') r+38 \gamma ]\\
&\quad -3 r (r^2-2 \gamma  f' r+2 \gamma ) h''\} h+5 r^3 (r^2-2 \gamma  f' r+2 \gamma ) h'^3\} f^2-2 h \{2 \gamma  \{[(r^2+4 \gamma ) f''-r (r^2+2 \gamma ) f^{(3)}] r^2\\
&\quad +6 r^2+16 \gamma +f' (2 \gamma  f'' r^3+3 r^3+10 \gamma  r)\} h^2+r \{\gamma  h' \{10 (r^2+2 \gamma )+r [f' (3 r^2-4 \gamma  f' r+10 \gamma )\\
&\quad +4 r (r^2+\gamma  f' r+2 \gamma ) f'']\}-r (r^2+2 \gamma ) (r^2-2 \gamma  f' r+2 \gamma ) h''\} h\\
&\quad +2 r^2 (r^2-2 \gamma  f' r+2 \gamma ) (r^2+\gamma  f' r+2 \gamma ) h'^2\} f+r (r^2+2 \gamma ) h^2 f' [r (r^2-2 \gamma  f' r+2 \gamma ) h'\\
&\quad +2 \gamma  h (f'' r^2+2)]\} \bigg\} / \bigg\{ 16 r^2 h^2 [2 \gamma -2 \gamma  (r f'+f)+r^2]^3 [h (2 \gamma -2 \gamma  f+r^2)-2 \gamma  r f h'] \bigg\}.
\end{split}
\label{eq:Veff}
\end{equation}
\end{widetext}
Numerical calculations are more convenient in terms of eq.~\eqref{LinEq} formulated in terms of $r$, rather than in terms or $r_\ast$ for two reasons. First, obtaining $V_\mathrm{eff}(r_\ast)$ requires to reconstruct the coordinate $r_\ast$ numerically for every change in the background or in the choice of $\gamma$, as can be seen from eq.~\eqref{eq:drast}. Second, $V_\mathrm{eff}$ contains up to third order derivatives of the background functions. These quantities are extremely inaccurate when they are obtained numerically.

\bibliography{bibnote}

\end{document}